\documentclass[11pt]{article} 

\usepackage{amsmath,amsthm,latexsym,amssymb,amsfonts,epsfig}

\newtheorem{Theorem}{Theorem}[section]
\newtheorem{Definition}{Definition}[section]
\newtheorem{Lemma}{Lemma}[section]

\newcommand{\be}{\begin{equation}}
\newcommand{\ee}{\end{equation}}
\newcommand{\ba}{\begin{eqnarray}}
\newcommand{\ea}{\end{eqnarray}}

\title{{\sf Properties of a smooth, dense, invariant domain}\\
 {\sf for singular potential Schr\"odinger operators}} 
\author{
{\sf T. Thiemann}$^1$\thanks{{\sf 
thomas.thiemann@gravity.fau.de}}\\
\\
{\sf $^1$ Inst. for Quantum Gravity, FAU Erlangen -- N\"urnberg,}\\
{\sf Staudtstr. 7, 91058 Erlangen, Germany}\\
}
\date{{\small\sf \today}}

\makeatletter
\@addtoreset{equation}{section}
\makeatother

\begin{document} 

\maketitle

{\sf

\begin{abstract}
Schr\"odinger operators often display singularities at the origin, the 
Coulomb problem in atomic physics or the various matter coupling terms 
in the Friedmann-Robertson-Walker problem being prominent examples. For 
various applications it would be desirable to have at one's disposal 
an explicit basis spanning a dense and invariant domain 
for such types of Schr\"odinger operators, for instance stationary 
perturbation theory or the Raleigh-Ritz method. 

Here we make the observation, that not only a such basis can indeed be provided 
but that in addition relevant matrix elements and inner products can be 
computed 
analytically in closed form, thus providing the required data e.g. for 
an analytical Gram-Schmid orthonormalisation.
\end{abstract}

\section{Introduction}
\label{s1}

The spectral theory of an (unbounded) operator starts with the question of 
its domain which should at least be dense in order that notions as 
adoint, symmetry and (essential) self-adjointness make sense \cite{1}.
Applications in physics include the important case of Schr\"odinger 
type operators on the Hilbert space $L_2(\mathbb{R}^n,d^nx)$ which consists
of a linear combination of a second derivative operator and a potential term.
If that potential term is just a polynomial in $x$ then a suitable domain
for that operator which in fact makes it symmetric is the Schwartz space
${\cal S}(\mathbb{R}^n)$ of smooth functions of rapid decrease in $x$.
If the potential term is not a polynomial but is polynomially bounded 
in $x$ then this domain is still valid. However, if the potential
is not polynomially bounded, then it displays singularities at some points
$x=x_k$ at which the potential term diverges. This is already 
the case for the hydrogen atom and more generally in molecular 
physics but there is a wide range of physical
systems where even more singular potentials occur as e.g. in quantum cosmology
\cite{2}. 

In this paper we consider the important class of potentials 
which are singular at one point which without loss of generality can be 
taken as the origin of the coordinate system. Possible generalisation to
a finite number of points will be discussed below. Furthermore, we 
require the potential to be polynomially bounded in both $||x||$ and 
$||x||^{-1}$. In this case as a domain the 
restriction of the Schwartz space to functions 
which vanish sufficiently fast at the origin comes into mind, for example by 
considering functions of the form $||x||^{2N} f,\; f\in {\cal S}(\mathbb{R}^n)$
if the polynomial growth of the potential at the origin is no worse than
$||x||^{-2(N-1)}$. However, such functions do not form an invariant 
domain, i.e. higher powers of the potential are ill defined on it. 
This makes such functions unsuitable for even standard stationary 
perturbation theory \cite{3}.

It transpires that an invariant domain must consist of smooth functions 
which are of rapid decrease in both $||x||$ and $||x||^{-1}$. While
it is easy to see that such functions exist \cite{4} to be useful 
in physical applications one would like to have at one's disposal a concrete
system of such functions which both have 1. a dense span and 2.
explicitly computable matrix elements between any operators which are 
polynomials in derivative operators, multiplication operators as well
as the inverse of the multiplication operator. In this paper, we introduce 
one possible such system.\\
\\
The architecture of this paper is as follows:\\
\\
In section 2 we propose the mentioned system of functions and prove that
it has the desired properties. The discussion will be limited to one 
degree of freedom $n=1$ but the generalisation to finitely many degrees
of freedom is straightforward.

In section 3 we consider as an application 
the generalisation to even more singular Hamiltonian
operators which in addition contain inverse powers of the derivative operator.
Such operators indeed occur in quantum cosmological perturbation theory
\cite{5} and here our results will be significantly weaker. 

In section 4 we conclude and discuss the generalisation to more than 
one singular point in the potential and to more than one degree of freedom

\section{The basis and its properties}
\label{s2}

We consider a Schr\"odinger operator in one dimension of the form 
\be \label{2.1}
H=\frac{P^2}{2m}+V(Q)
\ee
where $P,Q$ are operators on the Hilbert space ${\cal H}=L_2(\mathbb{R},dy)$ 
acting in the usual way 
\be \label{2.2}
(P\psi)(y)=i\hbar\; \frac{d\psi}{dy}(y),\;
(Q\psi)(y)=y\;\psi(y)
\ee
thus satisfying the canonical commutation relations 
$[P,Q]=i\hbar 1_{{\cal H}}$ on a common invariant domain such as the 
Schwartz space ${\cal S}(\mathbb{R})$ of smooth functions of rapid decrease
at infinity. Here the 
potential is allowed to be a polynomial of finite degree in both 
$Q$ and $Q^{-1}$ where however
\be \label{2.3}
Q^{-1}\psi)(y)=\frac{\psi(y)}{y}
\ee
is not defined on all of ${\cal S}(\mathbb{R})$. 

In this paper we consider the domain $D\subset {\cal S}(\mathbb{R})$ 
consisting of smooth functions 
of rapid decrease in both $y$ and $y^{-1}$, i.e. the functions that decay 
at both $y=\pm \infty$ and $y=0$ respectively faster than any 
power $y^n$ or $y^{-n}$ for any $n\in \mathbb{N}_0$. 
As is well known \cite{1}, an element of $D$ 
is given by 
\be \label{2.4}
b_0(y):=\exp(-\frac{y^2}{2 l^2}-\frac{L^2}{2y^2})
\ee
where $l,L$ are two, possibly different length scales that maybe suggested  
by the physical problem at hand (we are assuming $y$ to have dimension of 
length). Functions of the form (\ref{2.4}) without the Gaussian 
factor play an important role in the mollification of distributions 
\cite{1}. 

For instance we might consider a harmonic oscillator with 
frequency $\omega$ and thus length scale 
$l^2=\frac{\hbar}{m\omega}$ coupled to a singular potential of Coulomb type 
$V(Q)=V_0 \frac{L}{Q}$ where $V_0$ has dimension of energy. It will
be convenient to work with the dimensionfree variable 
\be \label{2.5}
x=\frac{y}{r}, \; \frac{r^2}{l^2}=\frac{L^2}{r^2}=a\;\;\Rightarrow\;\;
r^4=l^2\; L^2;\; a=\frac{L}{l}>0
\ee
so that 
\be \label{2.6}
b_0(x):=b_0(y=r x)=e^{-a\frac{x^2+x^{-2}}{2}}
\ee
We also define for $n\in\mathbb{Z}$
\be \label{2.7}
b_n(x)= x^n\; b_0(x)
\ee
\begin{Lemma} \label{la2.1}~\\
i. The span $D_0$ of the $b_n$ is contained in D.\\
ii. $D_0$ is an invariant domain for any polynomial in $Q, Q^{-1}, P$.\\
iii. $D_0$ is dense in $\cal H$.
\end{Lemma}
Proof:\\
i.\\ 
It is well known that $\lim_{x\to \pm \infty} x^n e^{-ax^2/2}=0$ for 
$a>0$
and $n\in \mathbb{Z}$ since a Gaussian is a Schwartz function.
Thus also $\lim_{x\to \pm \infty} x^{-n} 
e^{-ax^{-2}/2}=0$ for $a>0,\; n\in \mathbb{Z}$ by switching to 
$z=x^{-1}\to \pm \infty$. Finally 
$\lim_{x\to 0} e^{-a x^2/2}=\lim_{x\to \pm \infty} e^{-a x^{-2}/2}=1$.
Therefore $b_n$ is smooth also at $x=0, x=\pm \infty$.\\
ii. This follows easily from the formulae 
\be \label{2.8}
x\; b_n(x)=b_{n+1}(x),\;
x^{-1}\; b_n(x)=b_{n-1}(x),\;
\frac{d}{dx} b_n(x)=n b_{n-1}(x)-a(b_{n+1}-b_{n-3})(x)
\ee
and the commutation relations.\\
iii.\\
Suppose that $D_0$ is not dense. Then we find 
$0\not=f\in \overline{D_0}^\perp$ in the orthogonal complement of the 
closure of $D_0$ in ${\cal H}$. In particular, $<b_n,f>=0$ for all
$n\in \mathbb{N}_0$. It follows $<p_n,f'>=0$ for all $n\in \mathbb{N}_0$
where 
\be \label{2.9}
p_n(x)=x^n \; e^{-a x^2/2},\; f'(x)=e^{-a x^{-2}/2}\; f(x)
\ee
Note that $f'\in L_2(\mathbb{R},dx)$. 
However, the Gram-Schmidt orthonormalisation of the $p_n$ yields the 
Hermite functions which lie dense. It follows $f'=0$ as an $L_2$ function,
that is, $f'(x)=0$ a.e. wrt Lebesgue measure. Since $1\ge e^{-a x^{-2}/2}>0$
except at $x=0$ it follows $f=0$ a.e., that is, $f=0$ as an $L_2$ function
which is a contradiction.\\
$\Box$\\
\\
The $b_n$ can thus be used to provide an orthonormal basis of functions 
in $D_0$ by using Gram-Schmidt orthonormalisation 
which would be very helpful in practical applications 
such as stationary perturbation theory etc. The usefulness of the $b_n$
of course relies on whether we can analytically compute the  scalar products
$<b_m,b_n>$ as a numerical computation would be 
very cost intensive and would only be possible up to some finite 
$N\ge |m|, |n|$. Once these are known, we can compute 
matrix elements of any polynomial in $Q, Q^{-1}, P$ in  closed form 
using (\ref{2.8}). 

The perhaps surprising observation, demonstrated below, 
is that indeed $<b_m, b_n>$ can be computed explicitly. Note that trivially 
$<b_m,b_n>=0$ for $m+n$ odd because $b_m$ is 
real valued and $b_m(-x)=(-1)^m b_m(x)$. For $m+n$ even we have 
$<b_m,b_n>=I_{\frac{m+n}{2}}(a)$ where for $n\in \mathbb{Z}$
we define the integral
\be \label{2.10}
I_n(a)=\int_{\mathbb{R}}\; dx\; x^{2n}\; e^{-a[x^2+x^{-2}]}   
\ee
A further simplification consists in the observation that $I_n(a)$ 
for $n<0$ is already determined by the $I_n(a)$ for $n\ge 0$.
\begin{Lemma} \label{la2.2}~\\
For all $n\in \mathbb{Z}$ 
\be \label{2.11}
I_{-(n+1)}(a)=I_n(a)
\ee
\end{Lemma}
Proof:\\
The proof makes use of the fact that $b_0(x)=b_0(x^{-1})$ is invariant 
under taking reciprocal values $x\mapsto x^{-1}$ which is 
a diffeomorphism on $\mathbb{R}^+$ with Jacobean $x^{-2}$. We have 
\ba \label{2.12}
I_n(a) &=& 
2\;\int_0^\infty \; dx \; x^{2n}\; b_0(x)^2
\nonumber\\ 
&=&
2\;\int_0^\infty \; \frac{dx}{x^2} \; x^{-2n}\; b_0(x^{-1})^2
\nonumber\\ 
&=&
2\; \int_0^\infty \; dx \; x^{-2(n+1)}\; b_0(x)^2
\nonumber\\ 
&=& I_{-(n+1)}(a)
\ea
$\Box$\\
\\
The strategy to compute $I_n(a)$ explicitly is to 
consider the diffeomorphism 
\be \label{2.13}
z:\; \mathbb{R}^+\to \mathbb{R};\; x\mapsto x-x^{-1}
\ee
with Jacobean $1+x^{-2}$ and inversion 
\be \label{2.14}
x=\frac{z}{2}+\sqrt{1+[\frac{z}{2}]^2} 
\ee
The map (\ref{2.13}) is motivated by the identity 
\be \label{2.15}
x^2+x^{-2}=z^2+2\; \; \Rightarrow\; b_0^2(x)=e^{-2a}\; e^{-a z^2}
\ee
Thus, since by the above lemma we have $I_n=\frac{1}{2}[I_n+I_{-(n+1)}]$,
if we would manage to rewrite 
\be \label{2.16}
f_n(x):=\frac{x^{2n}+x^{-2(n+1)}}{1+x^{-2}}
\ee
as a polynomial 
in $z$ then the computation of $I_n(a)$ would be reduced to the evaluation 
of the moments of the Gaussian measure $b_0^2 \; dz$. 
\begin{Lemma} \label{la2.3}~\\
The following identity holds:
\be \label{2.17}
f_n(x)=(-1)^n\;[1+\sum_{l=1}^n\; (-1)^l \; q_l(x)];\; q_l(x)=x^{2l}+x^{-2l}
\ee
\end{Lemma}
Proof:\\
We have 
\be \label{2.18}
f_n(x) = x^{-2n}\; \frac{x^{4n+2}+1}{x^2+1} 
\ee
Furthermore we have the telescopic sum
\be \label{2.19}
(1+x^2)\sum_{l=0}^{2n}\; (-1)^l x^{2l}\; 
=\sum_{l=0}^{2n} \; (-1)^l\; x^{2l}
+\sum_{l=1}^{2n+1} \; (-1)^{l-1}\; x^{2l}
=x^{4n+2}+1
\ee
so that 
\be \label{2.20}
f_n(x)
=\sum_{l=0}^{2n}\; (-1)^l \; x^{2(l-n)}
=(-1)^n\;\sum_{l=-n}^n\; (-1)^l \; x^{2l}
=(-1)^n\;\{1+\sum_{l=1}^n\; (-1)^l \; [x^{2l}+x^{-2l}]\}
\ee
$\Box$\\
\\
The purpose of rewriting the integrand in terms of the $q_l$ is that these 
functions 
are now easily rewritten in terms of $z$ using (\ref{2.14}). For instance 
we have already seen that $q_1=z^2+2$.
\begin{Lemma} \label{la2.4} ~\\
We have the identity
\be \label{2.21}
q_l(x)=\sum_{k=0}^l\; [\frac{z}{2}]^{2k} a_k(l);\;
a_k(l)
=2 \sum_{r=l-k}^l
\left( \begin{array}{c} 2l \\ 2r \end{array} \right) \;
\left( \begin{array}{c} r \\ l-k \end{array} \right) \;
\ee
\end{Lemma}
Proof:\\
Let us use the abbreviations $u=\frac{z}{2},\; w=\sqrt{1+u^2}$, then
\be \label{2.22}
x=w+u,\; x^{-1}=w-u
\ee 
and 
\ba \label{2.23}
q_l(x) 
& =& \sum_{k=0}^{2l} \; 
\left( \begin{array}{c} 2l \\ k \end{array} \right) \;
w^k\; u^{2l-k}\;[1+(-1)^k] 
\nonumber\\
& =& 2 \sum_{k=0}^l \; 
\left( \begin{array}{c} 2l \\ 2k \end{array} \right) \;
w^{2k}\; u^{2(l-k)} 
\nonumber\\
& =& 2 \sum_{k=0}^l \; 
\left( \begin{array}{c} 2l \\ 2k \end{array} \right) \;
u^{2(l-k)} \sum_{r=0}^k 
\left( \begin{array}{c} k \\ r \end{array} \right) \;
u^{2(k-r)}
\nonumber\\
& =& 2 \sum_{r=0}^l \; u^{2r}
\sum_{k=l-r}^l
\left( \begin{array}{c} 2l \\ 2k \end{array} \right) \;
\left( \begin{array}{c} k \\ l-r \end{array} \right) \;
\ea
$\Box$\\
\\
We can now assemble (\ref{2.17}), (\ref{2.21}):
\begin{Lemma} \label{la2.5}~\\
We have the identity
\be \label{2.24}
f_n(x)=1+2 \sum_{k=1}^n \; [\frac{z}{2}]^{2k} \; b_k(n),\;
b_k(n)=\sum_{l=k}^n \; (-1)^{n+l}\; a_k(l)
\ee
\end{Lemma}
Proof:\\
We have 
\ba \label{2.25}
f_n(x) 
&=& (-1)^n\;[1+2\;
\sum_{l=1}^n\; (-1)^l \sum_{k=0}^l \; [\frac{z}{2}]^{2k}\; a_k(l)
\nonumber\\
&=& (-1)^n\;[1+2\;\sum_{k=0}^n\; [\frac{z}{2}]^{2k}
\sum_{l\ge 1,k}^n\; (-1)^l a_k(l)
\nonumber\\
&=& (-1)^n\;[(-1)^n+2\;\sum_{k=1}^n\; [\frac{z}{2}]^{2k}
\sum_{l=k}^n\; (-1)^l a_k(l)
\ea
where we used that $a_0(l)=1$ and 
\be \label{2.26}
\sum_{l=1}^n (-1)^l=\frac{(-1)^n-1}{2}
\ee
$\Box$\\
It remains to compute the moments of the given Gaussian measure.
\begin{Lemma} \label{l2.6} ~\\
We have 
\be \label{2.27}
\int_0^\infty \; dx\; [1+x^{-2}] (x-x^{-1})^{2k}\; b_0(x)^2
=\sqrt{\frac{\pi}{a}}\; e^{-2a}\;\frac{1}{a^k}
\;
\frac{(2k)!}{4^k\; k!}
\ee
\end{Lemma}
Proof:\\
The introduction of the variable $z=x-x^{-1}$ yields
the Gaussian integral
\be \label{2.28}
e^{-2a}\;\int_{-\infty}^\infty\;dz\; z^{2k}\; e^{-a z^2}
=\frac{e^{-2a}}{\sqrt{a} a^k}\;J_k,\;\;
J_k= \int_{\mathbb{R}}\; dz \; z^{2k} \;e^{-z^2}
\ee
for which an integration by parts yields the recursion 
\be \label{2.29}
J_k=\frac{2k-1}{2} J_{k-1}=\frac{(2k)!}{4^k\; k!} J_0
\ee
$\Box$\\
\\
We summarise our findings in the theorem:
\begin{Theorem} \label{th2.1}~\\
The integral (\ref{2.10}) can be computed in closed form and its value 
is explicitly given by
\be \label{2.30}
I_n(a)=\sqrt{\frac{\pi}{a}}\;e^{-2a}\;
[1+2\sum_{k=1}^n\; \frac{(2k)!}{4^{2k}\; k!\; a^k} \; b_k(n)]
\ee
where the coefficients $b_k(n)$ are defined by (\ref{2.24}) and (\ref{2.21})
(the sum over $k$ contributes only for $n\ge 1$).
\end{Theorem}
The value $I_n(a)$ involves a triple sum. It would be desirable to condense 
it using some identities for binomial coefficients in order to simplify 
its application in further computations such as the Gram-Schmidt 
orthonormalisation of the sequence of vectors 
\be \label{2.31}
b_0,\; b_1,\; b_{-1},\; b_2,\; b_{-2},\; ...,\; b_n,\; b_{-n},\;...
\ee
Unfortunately we could not yet find a way to resum the terms involved 
in $I_n(a)$. 

\section{Inclusion of inverse derivative operators}
\label{s3}

As mentioned in the introduction, there are physical systems whose 
Hamiltonian requires a dense and invariant domain for polynomials in 
$P,Q, P^{-1},Q^{-1}$. The operator $P^{-1}$ has the symmetric integral kernel
\be \label{3.1}
(P^{-1}\psi)(x)=\frac{1}{2i\hbar}\; \int_{\mathbb{R}}\; dy\; {\rm sgn}(x-y)
\psi(y)
\ee
but $P^{-1}$ is ill defined on the domain $D$ introduced in the previous 
section. For instance $P^{-1}b_0$ is no longer square integrable. It is easy
to see using the spectral theorem that the image of the Fourier transform 
of $D$ is an invariant domain for any polynomial in $Q, P, P^{-1}$ but 
now $Q^{-1}$ is ill defined for the same reason.

In what follows we will consider the class of Hamiltonian operators whose 
classical symbols $h(q,p)$ can
be written in the  form
\be \label{3.2}
h(q,p)=h_+(q,p)+h_-(q,p),\;
h_-(q,p)=\bar{F}(q)\;[\sum_{n=1}^N\; h_{-n}(q)\; p^{-n}]\; F(q)  
\ee
for any $N<\infty$.
Here $h_+(q,p)$ is a real valued 
polynomial in $p$ with coefficient functions that 
are polynomially bounded in both $q,q^{-1}$, $1/F(q)$ 
is polynomially bounded 
in both $q,q^{-1}$ and smooth everywhere except possibly at 
$q=0,\pm \infty$. Finally, $h_{-n}$ is a polynomial in $q$ only. Here we 
use the usual definition of polynomial boundedness \cite{1}:
\begin{Definition} \label{def3.1}~\\
A function $f: \mathbb{R}\to \mathbb{C}$ is said to be polynomially 
bounded in both $q,q^{-1}$ if there exist $M,N\in \mathbb{N}$ such that 
for all $q\in \mathbb{R}$ and some $C>0$
\be \label{3.0}
|f(q)|\le C\;[1+q^2]^M \;(1+q^{-2})^N
\ee
\end{Definition}
The form (\ref{3.2}) is 
certainly not the most general situation because it may not be possible 
to facor out a common function $|F|^2$ in order to achieve polynomiality 
of all the $h_{-n}$. In those more general situations, independent 
ideas of a kind similar to the one presented in the subsequent 
theorem work at least for some possible values of $n$ but the
general statement is then much weaker, hence we refrain from considering these
extensions of our results 
here which must be analysed in a case by case study.
\begin{Theorem} \label{th3.1}~\\
There exists a subspace $D_L(F)\subset D$ which is a domain for 
a symmetric ordering $H$ of (\ref{3.2}) and which is mapped by $H$ to 
$D$, the smooth functions of rapid decrease in both $x, x^{-1}$.
\end{Theorem}
Proof:\\
Let $H_+$ be any symmetric ordering of $h_+$ and consider the following 
symmetric ordering of $h_-$
\be \label{3.3}
H_-=\frac{1}{2}F(Q)^\dagger\;[\sum_{n=1}^N\; \{
h_{-n}(Q)\; P^{-n}+P^{-n} h_{-n}(Q)\}]\; F(Q)
\ee
Consider a general polynomial $r(q)$ of degree $M$ then 
\be \label{3.4}
[P^{-1}, r(Q)]=P^{-1}\;[r(Q),P]\;P^{-1}=-i\hbar\; P^{-1}\; r'(Q)\; P^{-1}
\ee
where $r'$ is the derivative of $r$.
By iterating (\ref{3.4}) it follows that we have an identity of the form 
\be \label{3.5}
P^{-n} h_{-n}(Q)=\sum_{k=0}^{M_n} \;c_{k,n}\; h_{-n}^{(k)}(Q) \; P^{-(n+k)}
\ee
where $M_n$ is the degree of $h_{-n}$ and $h_{-n}^{(k)}$ denotes the 
$k-$th derivatibe of $h_{-n}$ and $c_{k,n}\in\mathbb{C}$. 
Here it was important that $h_{-n}$
is indeed a polynomial as otherwise we could not have ordered all functions 
of $Q$ to the left of all functions of $P^{-1}$ in (\ref{3.5}) using
only finitely many terms. 

Let now 
\be \label{3.6}
L:={\rm max}_{n=1}^N(n+M_n)
\ee
Note that $L\ge N$. Now define $D_L(F)$ to be the space of functions
\be \label{3.7}
f\in D_L(F)\;\;\Leftrightarrow\;\; f(x)=\frac{g^{(L)}(x)}{F(x)},\; g\in D
\ee
It follows that $H D_L(F)\subset D$:  $F^{-1}$ is polynomially 
bounded in both $q,q^{-1}$ and smooth hence all its finite derivatives
exist except possibly at $q=0,\pm \infty$. But they exist also 
there when multiplied
by $g^{(L)}\in D$ which makes it a domain for $H_+$ which 
thus maps $D_L(F)$ to $D$.
Next $P^{-M} F$ maps $D_L(F)$ into $D$ because $P^{-M} g^{(L)}\propto
g^{(L-M)}$ which thus makes $D_L(F)$ a domain for $H_-$ which maps it to $D$
as well.\\
$\Box$\\
\\
Obviously, $D_L(F)$ is not an invariant domain for $H$ since we only have 
$H D_L(F)\subset D$. Moreover, it is not clear that $D_L(F)$ is dense. 
Suppose, however, that $f\in D$ and that  
\be \label{3.8}
<f,F^{-1} g^{(L)}>=(-1)^L\; <(\bar{F}^{-1} f)^{(L)},g>=0
\ee
for all $g\in D$.
Choosing $g=b_n$ we conclude that $(\bar{F}^{-1} f)^{(L)}=0$ using 
the results of the previous section.
Note that $(\bar{F}^{-1} f)^{(L)}\in {\cal H}$. 
It follows that 
\be \label{3.9}
f=\bar{F} r
\ee
where $r$ is an arbitrary polynomial of degree $L-1$. 
Since $F$ is nowhere vanishing except on a measure zero set (otherwise
$F^{-1}$ was not polynomially bounded), to have $\bar{F} r\in D$ 
requires either $F\in D$ or $r=0$. However, $F\in D$ means that 
in particular $F^{-1}$ is growing faster than any polynomial at infinity
which contradicts the assumption that 
$F^{-1}$ is polynomially bounded. Accordingly we must have 
$r=0$ and thus $f=0$. This shows that the orthogonal comeplement of 
$\overline{D_L(F)}$ at least has trivial intersection with the dense set 
$D$.        

To show that $D_L(F)$ is dense it would be sufficient to show that 
the space of the $r(x) g^{(L)}(x),\; g\in D$ is dense. 
Here, if it exists, $r(x)$ is 
a polynomial in $x,x^{-1}$
such that $(r F)^{-1}$ is a uniformly bounded function a.e. because then 
\be \label{3.10}
<\psi, F^{-1} g^{(L)}>=<\overline{(F r)^{-1}}\psi,r g^{(L)}>=0
\ee
for all $g$ would imply $\psi=0$. Now for $g\in D_0$ we certainly have 
$r g^{(L)}\in D_0$ and thus one could try to show that every $b_n$ is 
in the image of $g\mapsto r g^{(L)}, g\in D_0$. In view of 
the relation (\ref{2.8}) this is however not obvious. It is not even 
clear that every $b_m$ is in the finite linear span of the 
$b_{m}'$ so that this idea of proof is probably not very fruitful.

\section{Conclusion and Outlook}
\label{s4}

The results of section \ref{s2} reveal that the functions in $D_0$ are 
under similar analytical control as the Hermite functions. What is missing
is an explicit Gram-Schmidt orthogonalisation of the basis $b_n$ which we 
will leave for future research. We also showed that the results 
of section \ref{s2} are helpful for the study of Hamiltonian 
operators whose classical symbols are singular in both $q,p$ although 
our results here are significantly weaker.\\
\\
We conclude this paper by mentioning several 
extensions of the formalism developed:\\
1.\\
When there are a finite number $N$ of distinct singular points $x_k$ we could 
use the invariant 
domain $D_0$ for polynomilas in $P, Q, (Q-x_k \cdot 1_{{\cal H}})^{-1}$ 
defined as the finite linear span of the functions
\be \label{4.1}
b_{n_1,..,n_N}(x)=\prod_{k=1}^N\; b_{n_k}(x-x_k),\; n_k\in \mathbb{Z}
\ee
Far away from all the $x=x_k$ this is basically a Gaussian times the 
monomial $x^{n_1+..+n_N}$ and close to the singularity $x_k$ it behaves like 
$b_{n_k}(x-x_k)$ times a constant. Since we can write $\prod_k b_0(x-x_k)$
as a Gaussian in a variable $x'$ (which is defined by 
$\sum_k (x-x_k)^2=(x')^2+c$ where $c$ is a constant) times a function 
which is everywhere bounded and positive, except at $x=x_k$ where it vanishes,
by the argument of section \ref{s2} 
the span of the $b^{n_1,..,n_N}$ is dense. It is unlikely
that the computation
of the inner products of the (\ref{4.1}) is possible in closed form because 
the techniques of section \ref{s2} do not apply when $N\ge 2$ but perhaps 
an efficient approximation is possible based on a partition of $\mathbb{R}$ 
into neighbourhoods $I_k$ of $x_k$ where (\ref{4.1}) is basically 
$\propto b_{n_k}(x-x_k)$ and the complement $\mathbb{R}-\cup_k I_k$ 
where it behaves as $\propto b_{n_1+..+n_N}(x')$ in a suitably rescaled 
coordinate $x'=\sqrt{N} x$.\\
2.\\
If we consider more than one degree of freedom, i.e. several variables 
$x^a,\; a=1,..,D$ but a potential that still has 
only a singularity at the origin then we will proceed according to the 
symmetries of the 
Hamiltonian. At the extreme ends are a rotationally invariant Hamiltonian
and an Hamiltonian without any rotation symmetry. In the first case 
one will pass to polar coordinates, i.e. radius $r=||x||$ and angular 
coordinates $\phi$ on $S^{D-1}$ and consider functions of the form 
$f(r) Y(\phi)$ where $Y$ are eigenfunctions of the Laplacian on $S^{D-1}$ 
and $f\in D_0$. In the second, most general case we simply take the product 
vectors
\be \label{4.2}
b_{n_1,..,n_D}(x^1,..,x^D)=\prod_{a=1}^D\;b_{n_a}(x^a)
\ee
These functions still have analytically computable inner products 
and matrix elements for 
all polynomials in $P_a, Q_a, (Q_a)^{-1}$ for which it is a dense and 
inariant domain by a direction wise repetition of the argument 
made in section two.\\ 
3.\\
Yet more general is the case of both situations 1. and 2. 
which necessarily can have no longer full rotation symmetry and one can 
consider
\be \label{4.3}
b_{n_1^1,..,n_D^N}(x_1,..,x_D)=\prod_{a=1}^D\;\prod_{k=1}^N
b_{n_a^k}(x^a-x_k^a)
\ee
which combines the formulae (\ref{4.1}) and (\ref{4.2}) and defines an
invariant basis for polynomials in $P_a, Q^a,(Q^a)^{-1}$.\\
4.\\
In another direction, matrix elements of the form 
$<b_m,\; x^k\; |x|^l|\;b_n>$ are finite as one can see using the Cauchy 
Schwartz inequality and 
$||x^k\; |x|^l|\;b_n||^2=<b_n, x^{2(k+l)} b_n>$ but an analytical evaluation 
of its exact value cannot be assembelled from the formulae provided in this 
paper. More generally, the tools provided in this paper can be used to give 
estimates of matrix elements of any function of $x$ which is polynomially
bounded in both $x, x^{-1}$.\\
5.\\
Finally, a Gram-Schmidt orthonormalisation $o_n,\; n\in \mathbb{Z}$ of the 
sequence 
$b_0,\; b_1,\; b_{-1},\; b_2,\; b_{-2},..,\;b_n,\; b_{-n},..$ 
can be easily assembelled up to any finite $|n|$ 
using the formulae given in this paper but it would be helpful to have a 
closed formula of the 
expansion coefficients $<o_m,b_n>$ for any $m,n$ at one's disposal.\\
6.\\
In closing, we make the trivial remark, that the analytical results of this
paper straightforwardly generalise to $L_2(\mathbb{R}^n,\; d^n x)$
for $n<\infty$ and for any
Hamiltonian operators which are polynomials in the $p_a,q^a,\frac{1}{q^a},;\
a=1,..,n$. This applies in particular to constructive quantum field theory
where one regularises the interacting theory on finite lattices and then
uses renormalisation theory to remove the lattice (UV regulator)
before taking infinite volume limit (IR regulator).

}

\end{document}